\documentclass[reprint,
superscriptaddress,
amsmath,amssymb,
aps,
prl,
showpacs,
longbibliography
]{revtex4-1}
\usepackage{epsfig}
\usepackage[colorlinks=true,citecolor=blue,linkcolor=black,urlcolor=blue]{hyperref}

\begin{document}

\title{Strong higher-order resonant contributions to x-ray line polarization in hot plasmas}




\author{Chintan~Shah}\email{chintan@mpi-hd.mpg.de}
\affiliation{Physikalisches Institut der Universit\"at Heidelberg, 69120 Heidelberg, Germany}
\affiliation{Max-Planck-Institut f\"ur Kernphysik, Heidelberg, 69117 Heidelberg, Germany}

\author{Pedro~Amaro}
\altaffiliation{Present address: LIBPhys-UNL,~Departamento de F\'isica,~FCT-UNL,  P-2829-516, Caparica, Portugal}
\affiliation{Physikalisches Institut der Universit\"at Heidelberg, 69120 Heidelberg, Germany}

\author{Rene~Steinbr\"{u}gge}
\affiliation{Max-Planck-Institut f\"ur Kernphysik, Heidelberg, 69117 Heidelberg, Germany}

\author{Christian~Beilmann}
\altaffiliation{Present address: Karlsruhe Institute of Technology, 76131 Karlsruhe, Germany}
\affiliation{Physikalisches Institut der Universit\"at Heidelberg, 69120 Heidelberg, Germany}
\affiliation{Max-Planck-Institut f\"ur Kernphysik, Heidelberg, 69117 Heidelberg, Germany}

\author{Sven~Bernitt}
\affiliation{Max-Planck-Institut f\"ur Kernphysik, Heidelberg, 69117 Heidelberg, Germany}
\affiliation{Institut f\"ur Optik und Quantenelektronik, Friedrich-Schiller-Universit\"at, 07743 Jena, Germany}

\author{Stephan~Fritzsche}
\affiliation{Helmholtz-Institut Jena, 07743 Jena, Germany}
\affiliation{Theoretisch-Physikalisches Institut, Friedrich-Schiller-Universit\"at Jena, 07743 Jena, Germany}

\author{Andrey~Surzhykov}
\affiliation{Helmholtz-Institut Jena, 07743 Jena, Germany}

\author{Jos\'e~R.~{Crespo L\'opez-Urrutia}}
\affiliation{Max-Planck-Institut f\"ur Kernphysik, Heidelberg, 69117 Heidelberg, Germany}

\author{Stanislav~Tashenov}
\affiliation{Physikalisches Institut der Universit\"at Heidelberg, 69120 Heidelberg, Germany}

\date{\today}

\begin{abstract}
We studied angular distributions of x rays emitted in resonant recombination of highly charged iron and krypton ions, resolving dielectronic, trielectronic, and quadruelectronic channels. A tunable electron beam drove these processes, inducing x rays registered by two detectors mounted along and perpendicular to the beam axis. The measured emission asymmetries comprehensively benchmarked full-order atomic calculations. We conclude that accurate polarization diagnostics of hot plasmas can only be obtained under the premise of inclusion of higher-order processes that were neglected in earlier work.
\end{abstract}


\maketitle

The observation of x-ray polarization in emissions from the Crab Nebula~\cite{weisskopf1976, weisskopf1978}, with synchrotron radiation as its origin, fueled a strong interest in the astrophysics community for launching an x-ray polarimetry (XRP) mission~\cite{nature_news2013,costa2001,kamae2008,hiroyasu2010,chattopadhyay2014}. 
\emph{X-ray Imaging Polarimetry Explorer (XIPE)} has recently been selected as one of three candidates for the next medium-size satellite mission by the European Space Agency ESA. Its aim is studying the anisotropies of astrophysical plasmas which are found in the most extreme yet poorly understood sites in the Universe~\cite{krawczynski2011,soffitta2013}. 
Up to now anisotropic plasmas were found in active galactic nuclei~\cite{nayakshin2007,dovciak2004,dovciak2008}, pulsars~\cite{kallman2004}, gamma-ray bursts~\cite{coburn2003,gotz2013,gotz2014}, neutron stars~\cite{rees1975, weisskopf2006}, and solar flares~\cite{haug1972,haug1979,emslie2008}. They appear prominently also in the laboratory: in experiments with strong lasers~\cite{kieffer1992,kieffer1993}, magnetic cusps~\cite{iwamae2005}, \textit{z}-pinches~\cite{baranova2003}, and fusion devices~\cite{texter1986, fujimoto1997, baranova2003} such as tokamaks~\cite{inal1987,inal1989,fujimoto1996} and stellarators. The directionality of the electron--ion collisions leaves an imprint in the polarization of the plasma x rays~\cite{tseng1973, tashenov2011, eichler2002, tashenov2006, surzhykov2002, henderson1990}. Polarization measurements may therefore reveal the presence and orientation of particle beams, magnetic fields and, hence, provide information on the plasma heating and confinement mechanisms~\cite{texter1986, fujimoto1997, baranova2003} in instances where spatial resolution is insufficient.
In astrophysics, indeed, XRP is often the only technique for deriving information on the geometry of angularly unresolved sources~\cite{lei1997,krawczynski2011}.

Ideally, the XRP data should be analyzed with detailed knowledge of the atomic polarization mechanisms. Until now, however, no or little experimental information is available for most atomic processes. 
%
%
Only very few studies of x-ray polarization and angular distributions have been performed for astrophysically relevant ions using electron beam ion traps (EBITs)~\cite{beiersdorfer1996, beiersdorfer1997, shlyaptseva1998, shlyaptseva1999, beiersdorfer1999, beiersdorfer2001, nakamura2001, robbins2004, robbins2006}. 
Other studies with EBITs~\cite{takacs1996, hu2012, hu2014, ralchenko2013}, electron accelerators~\cite{tashenov2011,martin2012,tashenov2013} and storage ring~\cite{weber2010,tashenov2006} have focused on heavy ionic systems. 
Moreover, the photoelectric gas polarimeter~\cite{costa2001,bellazzini2013} of the {XIPE} mission will not resolve individual x-ray transitions. This limitation is rather general, even if high resolution detectors were used, since Doppler shifts will most likely blur the signal. Thus, the polarization signal will contain contributions from many transitions and continuum radiation due to bremsstrahlung and recombination of plasma electrons. This raises the needs for atomic data even higher by demanding a systematic knowledge of polarization of all contributing channels. 
The directional anisotropies and polarizations of the continuum radiation due to bremsstrahlung and radiative recombination are reasonably well understood~\cite{tashenov2006,tashenov2011,tashenov2013,tashenov2014}. In contrast, these properties were not sufficiently studied for bound--bound and resonant free--bound transitions producing strong x-ray lines.

\begin{figure*}
\includegraphics[width=\textwidth]{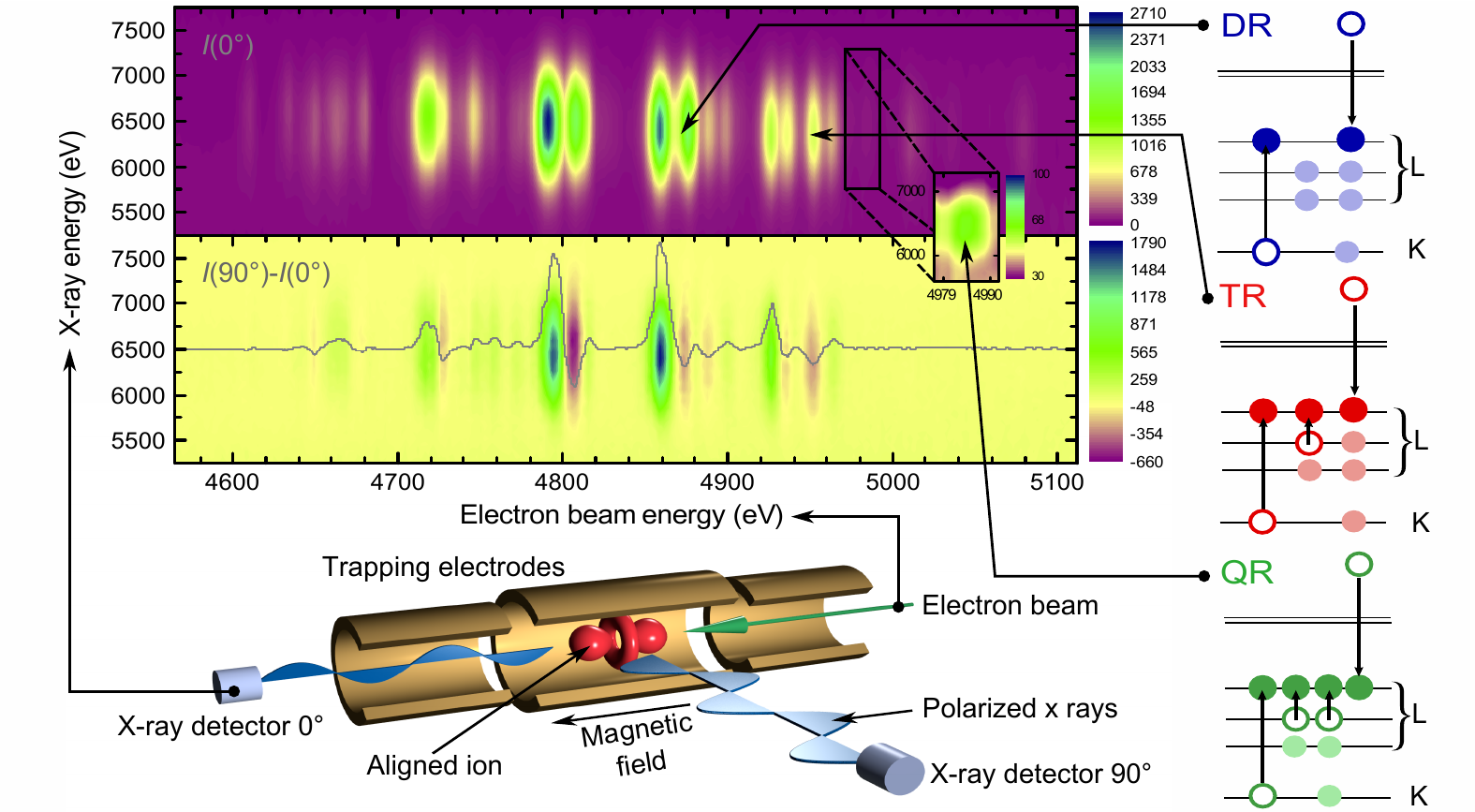}
\caption{Experimental setup: Ions are produced and trapped within a monoenergetic electron beam. 
Its energy was tuned into the recombination resonances, and the subsequently emitted x rays were observed by two germanium detectors at $0^\circ$ and $90^\circ$ with respect to the beam axis. 
Upper two-dimensional (2D) plot: Intensity of x rays along beam axis as a function of electron and x-ray energy. 
Lower 2D plot: Difference in intensity between the two $0^\circ$ and $90^\circ$ detectors. 
Solid line: Projection of this difference. 
Some individual resonances are marked exemplarily for the different resonant channels: DR, TR, and QR.
\label{fig:setup}}
\end{figure*}

In this letter we present a complete measurement of emission anisotropies of iron and krypton $K\alpha$ x-ray lines, and model their polarization. 
The polarization of x-ray lines from ions plays a strong role in respectively astrophysical and tokamak plasmas~\cite{haug1981, sazonov2002,bitter1993,widmann1995,zhuravleva2010,marin2014,marin2014a}. 
%
%
Using an EBIT we induced x-ray transitions by electron impact, resolving \emph{K}-shell dielectronic (DR)~\cite{massey1942,burgess1964}, trielectronic (TR), and quadruelectronic (QR)~\cite{beilmann2011} recombination channels in ions of interest. While DR was known to dominate the $K\alpha$ x-ray line formation, the latter two (higher-order) channels were previously considered insignificant for plasma modeling. Contrary to expectations, we found that they strongly affect the polarization of $K\alpha$ x rays emitted by plasmas. 
This result calls for more systematic experimental investigations of XRP properties of atomic transitions, which are now known almost exclusively from theory, that is in part known to deviate from measurements~\cite{nakamura2001,robbins2004,beiersdorfer2003,beiersdorfer2015}. 

In the first step of the resonant recombination, a free electron is captured under the excitation of one or more bound electrons
, producing an aligned intermediate excited state. In the second step, radiative relaxation yields an ion in a charge state lower by a unit. In \emph{KLL} resonances, the bound electron is excited from the \emph{K}- shell to the \emph{L}- shell by recombination of a free electron into the \emph{L}- shell. 
Resonant recombination is strong in astrophysical plasmas~\cite{dubau1980,beilmann2013}-- it cools them and strongly affects their charge balance~\cite{lagattuta1997,foord2000,orban2010,beiersdorfer2015}.
Recombination rates, required for plasma modeling, are extracted from laboratory measurements of DR x-ray yields. Such experiments rely on theory predicting the x-ray emission asymmetries~\cite{beiersdorfer1992,fuchs1998,radtke2000,beilmann2009,yao2010,hu2013,ali2011,mahmood2012}. DR is sensitive to the Breit interaction~\cite{fritzsche2009,hu2012,nakamura2008,beilmann_pa2013,beilmann2013,shah2015} in both the angular distribution and the linear polarization of the emitted x rays~\cite{fritzsche2009}. These properties were measured only for a few resonances in heavy ions~\cite{ma2003,hu2012,ralchenko2013,joerg2015}, and no systematic studies of the alignment in DR, TR or QR were reported.

\begin{figure*}[t]
\includegraphics[width=\textwidth]{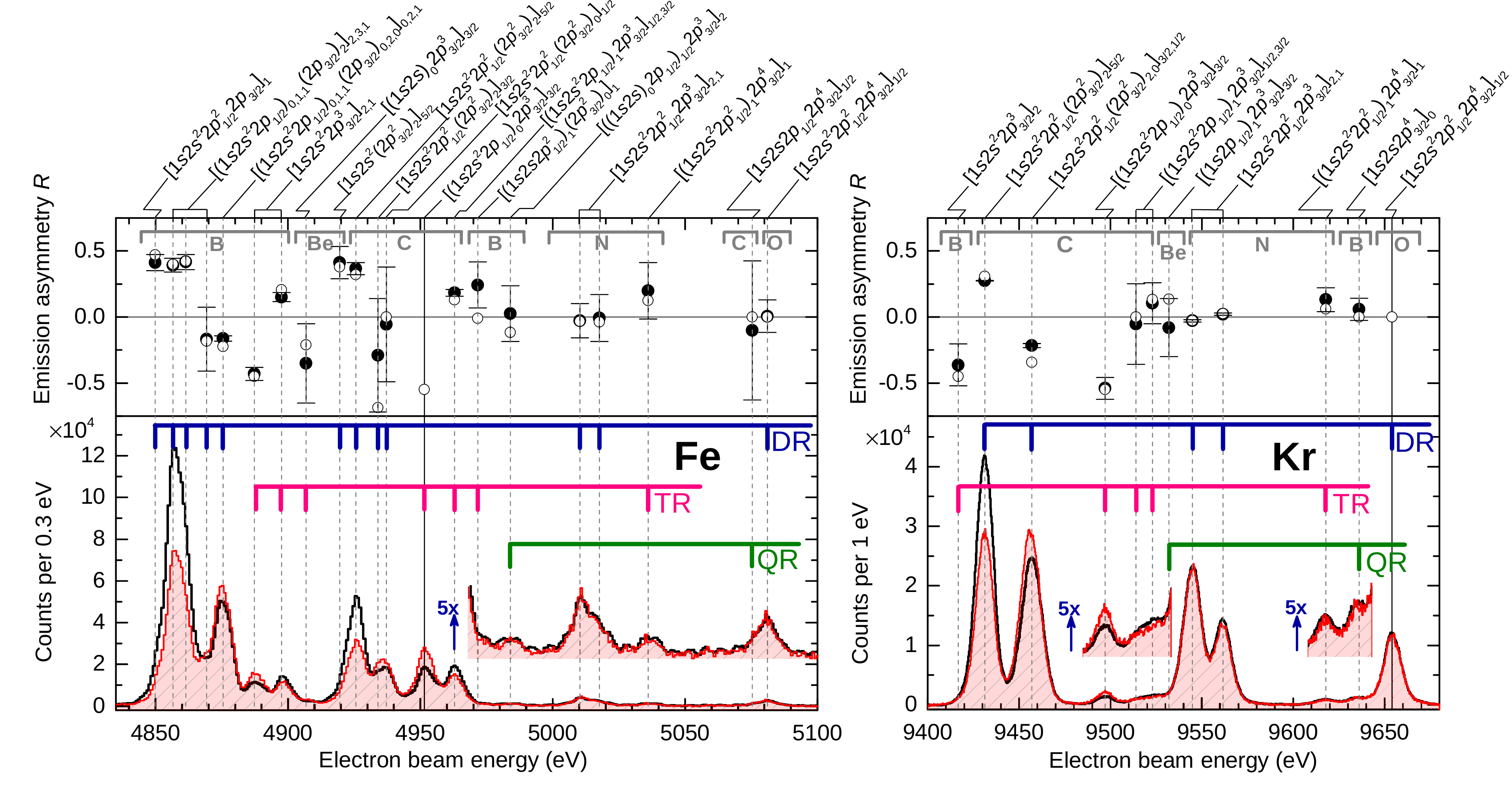}
\caption{X-ray intensities measured as a function of collision energy for iron (lower left panel) and krypton ions (lower right panel). The solid lines and the solid areas correspond respectively to the x rays observed side on and head on to the beam axis. 
DR, TR and QR resonances are indicated by the gray atomic symbols of the initial ion on the upper panels and by the configurations of the intermediate states. The extracted emission asymmetries \emph{R} (solid circles) for the individual resonances and the corresponding \textsc{FAC} predictions (open circles) are shown on the upper panels. Solid vertical lines indicate the resonances used for the intensity normalization.
\label{fig:spectra}}
\end{figure*}

The experiments were performed at the FLASH-EBIT~\cite{epp2007} 
with trapped ions in the He-like through O-like isoelectronic sequences produced by successive electron-impact ionization. 
The negative space charge of the monoenergetic electron beam traps positive ions radially, while the axial confinement results from the electrostatic potentials applied to the drift tubes surrounding the beam, see Fig.~\ref{fig:setup}. 
The x rays emitted by trapped ions were observed by two germanium detectors along and perpendicular to the electron beam axis while the electron energy was scanned over the range of the \emph{KLL} resonances.
The upper data inset of Fig.~\ref{fig:setup} shows the intensity of $K\alpha$ x rays observed by the first detector as a function of the electron and the x-ray energies. The intensity enhancements at given energies correspond to the recombination resonances. Moreover, the difference between the x-ray intensities observed by the two detectors, shown in the lower data inset, indicates anisotropic x-ray emission. 

The intensity of the $K\alpha$ transitions is shown in the lower panels of Fig.~\ref{fig:spectra} as a function of the electron collision energy. 
The x-ray background arising from the radiative recombination and the ambient radiation was subtracted.
Using the evaporative cooling technique~\cite{beilmann2010} we achieved the collision energy resolution of 6.5~eV full width at half maximum (FWHM) for iron and 11.5~eV for krypton, higher than in any previous EBIT experiment, hereby uncovering a large number of DR, TR, and QR resonances. We identified them using calculations performed with the Flexible Atomic Code (\textsc{FAC})~\cite{gu2008}. 
We included extended sets of configurations with the full configuration interaction and mixing between the states~\cite{beilmann_pa2013,yao2010}. 
The resonant electron capture is described using the distorted-wave formalism~\cite{gu2008}, and the full relativistic form of the electron--electron interaction, including the Breit interaction term, is taken into account. 
We fitted the experimental data with Gaussian profiles treating the line centroids and their intensities as free parameters and the resonance width fixed to the collision energy resolution. 
Reliable fits of blended resonances were possible due to very high counting statistics. 
The spectrum was calibrated with two strong and well isolated lines using their theoretical energies.

We corrected the data for the solid angles of the detectors; their ratio in the krypton measurement $\Omega_{0^{\circ}}/\Omega_{90^{\circ}}= 0.1068 \pm 0.0007$ was obtained using the isotropic decay of the $[1s2s^{2}2p^{2}_{1/2}2p^{4}_{3/2}]_{J=1/2}$ state. 
The isotropic iron lines had low intensities, and we used instead the TR transition exciting the $[(1s2s^2 2p_{1/2})_0 2p^3_{3/2} ]_{J=3/2}$ resonance. 
The radiative decay of this state has a theoretical intensity ratio of $I(0^{\circ})/I(90^{\circ})=1.55$ which is not dependent on the Breit interaction~\cite{gail1998}. 
The measured ratio $\Omega_{0^{\circ}}/\Omega_{90^{\circ}}= 0.082 \pm 0.003$ was different due to small changes of the setup.

From Fig.~\ref{fig:spectra} it is apparent that most of the observed resonant transitions are anisotropic. We quantify the emission asymmetries by the ratios
\begin{equation}
R \equiv \frac{I(90^{\circ})-I(0^{\circ})}{I(90^{\circ})}.
\label{eq:P90}
\end{equation}
The effects of finite detector solid angles and ion trap extension as well as cyclotron motion of the electrons~\cite{gu1999, beiersdorfer2001} were estimated to reduce the measured asymmetries only within the present statistical error bars.

\begin{figure}
\includegraphics[width=\columnwidth]{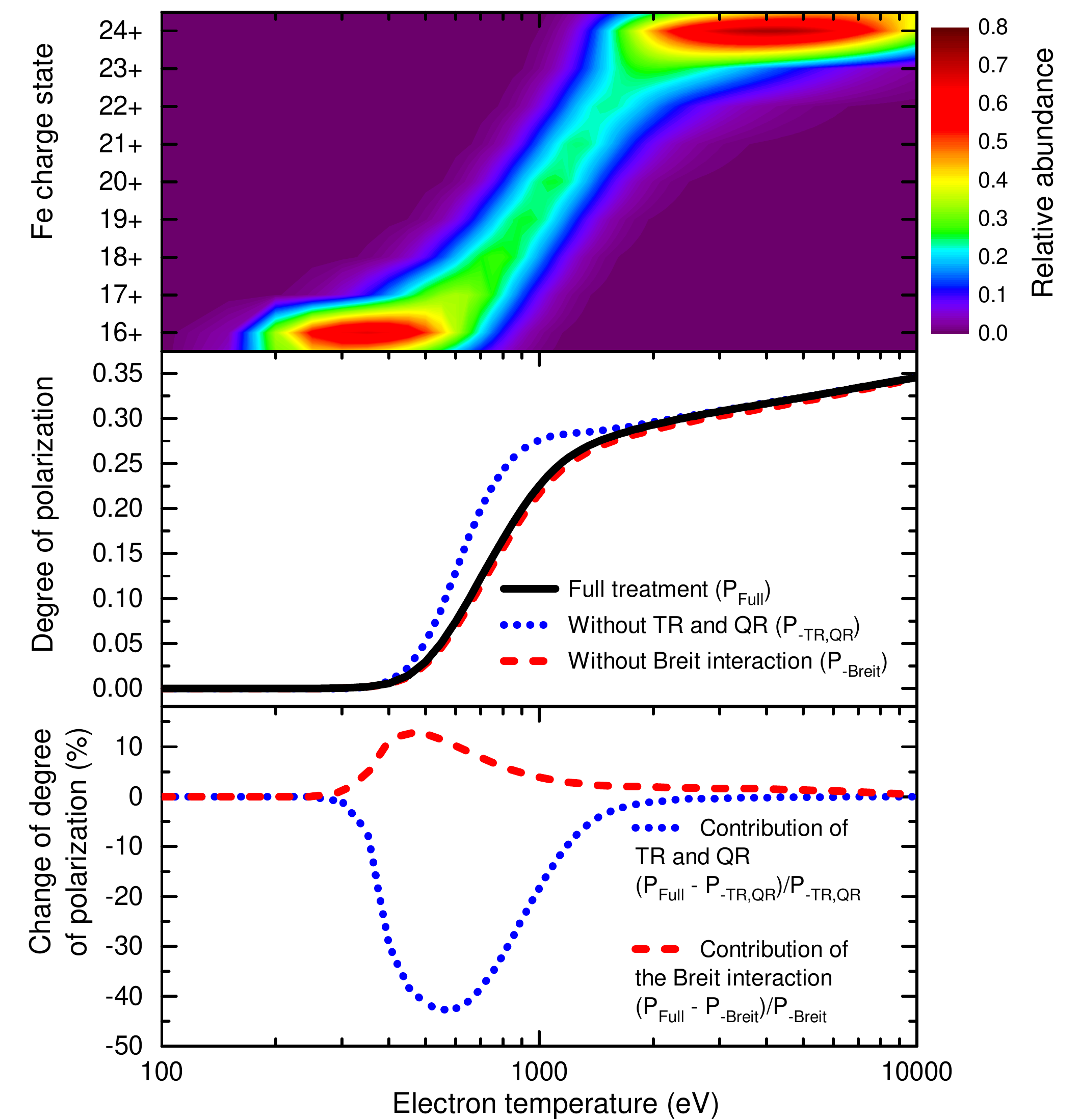}
\caption{Charge state distribution in an optically thin iron plasma (upper panel) and polarization of iron K$\alpha$ x-ray line $P(T)$ (middle panel) calculated with \textsc{FAC} as functions of temperature. The contributions of hitherto neglected higher-order transitions and the Breit interaction to the total polarization are shown in the middle and lower panels.
\label{fig:plasma}}
\end{figure}

Emission asymmetries result from nonstatistical populations of magnetic sublevels in the intermediate excited states. 
Since both electrons and ions are unpolarized, the population distribution can be described by a finite set of alignment parameters $\mathcal{A}_{k}$ (where \textit{k} is even) that define the angular distribution and polarization of the emitted x rays~\cite{chen1995,surzhykov2002}. 
%
%
In the electric dipole ($E1$) approximation these two x-ray properties are not affected by the alignment parameters with $k > 2$. 
Since all the observed radiative transitions are mainly of $E1$ type (other multipoles are smaller by five orders of magnitude), we restrict the further analysis to effects related to the alignment parameter $\mathcal{A}_{2}$~\cite{balashovbook, beiersdorfer1996}.

The angular distribution of the $E1$ transition from the intermediate to the final state is~\cite{balashovbook, surzhykov2006}
\begin{equation}
I(\theta) \propto 1 + \mathcal{A}_{2}  \alpha_{2} \left(1-\frac{3}{2}\sin^2\theta\right),
\label{eq:angdist}
\end{equation}
where $\alpha_{2}$ is an intrinsic anisotropy parameter determined by the total angular momenta of the intermediate and the final state, and $\theta$ is the emission angle with respect to the collision axis~\cite{balashovbook, beiersdorfer1996}. 
According to (\ref{eq:angdist}), in this experiment we have determined the product
\begin{equation}
\mathcal{A}_{2}\alpha_{2} = -\frac{2R}{3-R}.
\label{eq:a2r}
\end{equation}
Within the leading $E1$ approximation the same product defines the degree of linear polarization of x rays, emitted at the angle $\theta$ with respect to the collision axis, as~\cite{chen1995}
\begin{equation}
P(\theta) \equiv \frac{I_{\parallel}-I_{\perp}}{I_{\parallel}+I_{\perp}}
= \frac{R\sin^2\theta}{1 + R\cos^2\theta}.\label{eq:p2}
\end{equation}
The linear polarization is described by the intensities of the x rays $I_{\parallel}$ and $I_{\perp}$ polarized along  and perpendicular to the plane containing the collision and x-ray emission axes.
This equation indicates that the linear polarization of characteristic radiation emitted perpendicular to the collision axis coincides with the emission asymmetry: $P(90^{\circ}) = R$. 

We compare in the upper panels of Fig.~\ref{fig:spectra} the experimental emission asymmetries $R$ with \textsc{FAC} and \textsc{RATIP}~\cite{fritzsche2012b} predictions, in which account was taken of resonance strengths and unresolved radiative transitions into different final states. Both calculations produced identical results. 
With few exceptions, the agreement with the experiment is excellent, even for the higher-order \emph{KLL} resonances, known only since a few years ago~\cite{beilmann2011,beilmann2013,beilmann_pa2013}. 
Moreover, due to this systematic agreement we also conclude that the theoretical prediction for the emission asymmetry of TR resonance used for the solid angle normalization is also experimentally confirmed.

The photoelectric~\cite{costa2001} and Compton polarimeters~\cite{weber2015}, to be used in plasma polarization diagnostics, cannot resolve individual resonances. 
Thus, we calculate the cumulative effect of all \textit{KLL} transitions on the polarization of x rays as a function of temperature $T$. 
An optically thin plasma is assumed, which is a good approximation for solar flares and tokamaks. 
The temperature dependence arises due to abundances of ions $C^{i}(T)$ with the charge state $i$, as shown in the upper panel of Fig.~\ref{fig:plasma}. 
They are calculated with the inclusion of~\textit{KLL} higher-order channels using \textsc{FAC}, following the method presented by Gu~\cite{gu2003rr,gu2003dr}. 
We observed that not only low-energy TR~\cite{schnell2003,orban2010}, but also \emph{KLL} TR and QR significantly enhance the total recombination rates, thereby modifying the charge balance of the plasmas. 
%
The $C^{i}(T)$ with the resonance strength $S^i_n$ of individual transitions and $R^i_n$, where $n$ being the resonance number, define the maximum polarization $P(T)$ as
\begin{equation}
P(T) = \frac{ \sum\limits_{i,n} C^{i}(T) S^i_n R^i_n /(3-R^i_n)}{
 \sum\limits_{i,n} C^{i}(T) S^i_n  /(3-R^i_n)}\, .
\end{equation}
The maximum polarization corresponds to maximally anisotropic plasma electrons, i.~e., all electrons propagating perpendicularly to the observation direction. 
As shown for iron in the two lower panels of Fig.~\ref{fig:plasma}, the contributions of hitherto neglected higher-order channels reduce the polarization of $K\alpha$ line in the temperature range 500--1500~eV, which typically appear in solar flares~\cite{aschwanden1997,doschek1971,phillips2006a}. 
At these temperatures B-like through Ne-like iron ions dominate. 
Therefore, TR transitions in B- and C-like ions, populating the states $[1s2s^22p^3_{3/2}]_{J=2}$ and $[(1s2s^2 2p_{1/2})_02p^3_{3/2}]_{J=3/2}$,  are among the strongest, and they are responsible for this effect. 
We also point out the importance of the Breit interaction for accurate plasma polarization diagnostics. 
Verifying that these observations are not only iron specific, we obtained a similar reduction due to higher order in krypton at temperatures of over 2000~eV, which commonly occurs in tokamaks. 
In this temperature range, we note that the resonant capture process cross sections are a few orders of magnitude higher than other atomic processes also leading to the polarized emission of the $K\alpha$ x-ray line, namely electron-impact excitation in highly charged ions.

Our straightforward, but sensitive, experimental technique reduces greatly the time required for such comprehensive measurements compared to direct polarization measurements which require a dedicated x-ray polarimeter~\cite{weber2015,shah2015}. 
The present method is simple to implement and can be applicable to all elements of astrophysical and fusion research interest. 
Alignment resulting from many collisional processes, such as radiative recombination, as well as electron-impact excitation and ionization, can be studied with it. 
Ubiquitous but hitherto unrecognized higher-order channels~\cite{beilmann2011}, that can be as strong as the dielectronic process, should play an important role in the charge balance determination~\cite{beiersdorfer2015}. 
Our results pinpoint that, a too simplified approach, neglecting higher-order resonances and relativistic effects in the calculations, can significantly overestimate the plasma polarization. 
A systematic understanding atomic polarization requires both theoretical and experimental knowledge of previously neglected higher-order effects, as shown in this work for common types of astrophysical and laboratory plasmas.

\begin{acknowledgments}
This work was supported by the Deutsche Forschungsgemeinschaft (DFG) within the Emmy Noether program under Contract No. TA 740 1-1 and by the Bundesministerium f\"ur Bildung und Forschung (BMBF) under Contract No. 05K13VH2. We thank Dr. Zolt\'an Harman for valuable discussions on this work.
\end{acknowledgments}


%

\end{document}